\documentclass
[%
reprint,
superscriptaddress,
 amsmath,amssymb,
 aps,
 prl
]
{revtex4-2}
\usepackage{amsthm}
\usepackage{graphicx}
\usepackage{hyperref}
\usepackage[mathlines]{lineno}
\usepackage{xcolor}
\usepackage[utf8]{inputenc}
\usepackage[english]{babel}
\usepackage{soul}
\begin{document}
\title{Parameter free determination of optimum time delay}
\author{Thiago Lima Prado}
\affiliation{Departamento de F\'isica, Universidade Federal do Paran\'a, 81531-980 Curitiba, PR, Brazil.}
\author{Vandertone Santos Machado}
\affiliation{Departamento de F\'isica, Universidade Federal do Paran\'a, 81531-980 Curitiba, PR, Brazil.}
\author{Gilberto Corso}
\affiliation{Departamento de Biof\'isica e Farmacologia, Universidade Federal do Rio Grande do Norte,   59078-970 Natal, RN, Brazil}
\author{Gustavo Zampier dos Santos Lima}
\affiliation{\small Escola de Ci\^encias e Tecnologia, Universidade Federal do Rio Grande do Norte,   59078-970 Natal, RN, Brazil}
\author{Sergio Roberto Lopes}\email{lopes@fisica.ufpr.br}
\affiliation{Departamento de F\'isica, Universidade Federal do Paran\'a, 81531-980 Curitiba, PR, Brazil.}
\date{\today}

\begin{abstract}
 We show that the same  maximum entropy principle applied to recurrence microstates configures a new way to properly compute the time delay necessary to correctly sample a data set. The new method retrieves results obtained using traditional methods with the advantage of being independent of any free parameter. Since all parameters are automatically set, the method is suitable for use in artificial (computational) intelligence algorithms, recovering correct information embedded in time series, and  rationalizing the process of data acquisition since only relevant data must be collected. 
\end{abstract}

\maketitle

One of the essence of science is the recovery of meaningful information from (noisy) data. For long, modern data analysis has abandoned the assumptions of linearity and stationarity which marked conventional analysis and the assumption of non-linear properties brought great challenges since mathematical methods involved are far more complex \cite{kantz_2004,priestley_1988}. In this scenario and since experimental and simulated data are (primordially) continuous in time and/or space, the correct translation from such a continuous information to discrete data is crucial in preserving variable dependencies and meaning of the physical system, having a strong impact on the quality of the collect data. 

In the last decades, concepts from information theory have been used in order to extract the maximum of information embedded in data \cite{von_2011}. In special for a one dimensional time series, a scalar data set, the time interval that leads to minimums in the time delayed mutual information function provides a good systematic criterion for successfully build a discrete time series and to extract reliable information from continuous data set and/or to reconstruct phase portraits \cite{fraser_1986,kantz_2004}. It makes sense to use the first minimum of the time delayed mutual information function since this is the time lag where a future state of the time series adds maximal information to the present one and the redundancy is least \cite{fraser_1986,kantz_2004}. Since then, several data methodologies have used time series sampled in time intervals of the order of the first minimum of the time delayed mutual information \cite{von_2011}. However, we do not find (to our knowledge) physical principles to believe that always has to be a (first) minimum of the time delayed mutual information function. Another important question is the fact that higher time resolution must be provided to get the optimal time delay using the mutual information concept.  

Here,  we move a step forward in this field, building a method of automatically defined parameters to calculate the ideal time delay $\tau$ for data analysis based on the repetitive use of the  maximum entropy criterion \cite{jaynes_1957}, configuring a parameter free method. The automated capture of the ideal time delay allows artificial intelligence algorithms to develop the necessary data analysis, leading to a fast and reliable continuous delivery of results. In this way, we introduce a basic physical principle as the only criterion to be evaluated in the artificial intelligence algorithm for time series information recovery, rationalizing computing power to collect and to process only the relevant data. 

To construct an appropriated information entropy \cite{shannon_1948}, we make use of the recurrence entropy. This quantity is  built based on the diversity of \textit{recurrence microstates} (\textbf{RM}) \cite{corso_2018}, computed using all possible recurrences  between two randomly chose segments of time series of size  $N=1,\, 2,\, \cdots $ embedded in a time series of size $K$. The traditional representation of recurrence of a time series compares just two values, $x_i$ and $x_j$ of the time series, in this case $N=1$ and we have  \cite{eckmann_1987}
\begin{equation}
    \mathbf{R}_{ij}(\varepsilon)=\theta (\varepsilon-|{x_{i}}-{x_{j}}|),
    \label{recurrece} 
\end{equation}
where $\theta$ is the Heaviside step function and $\varepsilon$ is the recurrence vicinity threshold, a free parameter. For $|{x_{i}}-{x_{j}}| < \varepsilon$ in Eq. (\ref{recurrece}), ${x_{i}}$ and ${x_{j}}$ are said do be recurrent to each other and $\mathbf{R}_{ij}=1$ ($\mathbf{R}_{ij}=0$ otherwise). In this case, the two possible  \textbf{RM} are 1-digit binary values $\{0\}$ and $\{1\}$ \cite{corso_2018}. Considering now the \textbf{RM} of two sequences of two consecutive values, $\mathbf{x_i} = x_i, x_{i+1}$ and  $\mathbf{x_j} = x_j, x_{j+1}$ randomly chose in the time series, we get a set of $16$ possible \textbf{RM}, identified as $4$-digit binary values spanning from $\{0,0,0,0\}$ to $\{1,1,1,1\}$ \cite{corso_2018}. Analogously, for two sequences of three consecutive values $\mathbf{x_i} = x_i, x_{i+1},x_{i+2}$,  and $\mathbf{x_j} = x_j, x_{j+1},x_{j+2}$, we get $512$ possible \textbf{RM} composed of $9$-digit binary values. In general, considering two sequences of $N$ consecutive values, we get $2^{N^2}$, possible \textbf{RM}, each one composed of $N^2$-digit binary value.

Considering a sufficient large set of \textbf{RM} randomly selected in a time series, the density of recurrence points and  traditionally called recurrence rate ($RR$) \cite{marwan_2007}, can now be written as a function of the \textbf{RM} as
\begin{equation}
RR(\varepsilon)=\frac{1}{N^2} \sum_{w=1}^{N^2}{w\mathsf{P}_w\left[\mathbf{RM}(\varepsilon)\right]},
\label{rr}
\end{equation}
where $w$ is the number of  recurrent elements of a  $\mathbf{RM}$ (the number of ``ones" in a \textbf{RM}) and $\mathsf{P}_w$ is the probability to find a $w$-recurrent \textbf{RM}. Other recurrence quantifiers can also be written as a function of \textbf{RM} since they are functions of $RR$. 

The relation between the time delayed mutual information function ($M$) and the recurrence rate can be written approximately as a function of the recurrence rate, considering the case $N=1$ by using the second order Renyi entropy ($\hat{H_{2}}$) as \cite{Marco_2003,marwan_2007}:
\begin{equation}
{M}(\tau,\varepsilon))=-\hat{H}_{2}(x(t),x(t+\tau))+2\hat{H}_{2}(x(t),x(t)),
\label{mutual_info1}
\end{equation}
where
\begin{eqnarray}
\hat{H}_{2}(&x(t)&, x(t+\tau);\varepsilon)=\nonumber \\ 
&-&\ln \bigg [ \frac{\sum_{i,j}^{K}{\mathrm{\mathbf{R}}_{ij}(\varepsilon)}(x(t)){\mathrm{\mathbf{R}}_{ij}(\varepsilon)}(x(t+\tau))}{K^{2}}\bigg]. \label{mutual_info2}
\end{eqnarray}
Eq. \ref{mutual_info1} for the time delayed mutual information can also be written directly as a function of the \textbf{RM} themselves as 
\begin{equation}
\begin{split}
{M}(\tau,\varepsilon)=\ln \frac{1}{N^2}\bigg(\sum_{w=1}^{N^2}{w\mathsf{P}_w\left[\mathbf{RM}_\tau(\varepsilon) \circ \mathbf{RM}(\varepsilon)\right]}\bigg)\\-2\ln \frac{1}{N^2}\bigg(\sum_{w=1}^{N^2}{w\mathsf{P}_w\left[\mathbf{RM}(\varepsilon)\right]}\bigg),
\end{split}
\label{mutual_info}
\end{equation}
where $\mathbf{RM}_\tau(\varepsilon)$ is recurrence microstates computed for the $\tau$-delayed time series $(x_{0+\tau}, x_{1+\tau}, x_{2+\tau}, \cdots x_{K+\tau})$ and $[\circ]$ denotes the Hadamard Product.

To compute an entropy based on \textbf{RM}, we  sample a large number of $N$-value-segments (we have used  $1\%$ of all possible recurrence  microstates of the $K\times K$ binary matrix and we have focused on the simplest case  $N=2$) embedded in a $\tau$-delayed time series.  Calling $P_{w}$ the probability to obtain each possible \textbf{RM}, and making use of the Shannon entropy \cite{shannon_1948}, we obtain the entropy of \textbf{RM} \cite{corso_2018}.
\begin{equation}
S(\varepsilon,\tau)=-\sum_{w=1}^{2^{N^2}}P_{w}\left[\mathbf{RM}_\tau(\varepsilon)\right]\ln\{ \mathsf{P}_w\left[\mathbf{RM}_\tau(\varepsilon)\right]\}.
    \label{entropy}
\end{equation}
Usually, $\varepsilon$ and $\tau$ are the  recurrence vicinity parameter and the time delay, free parameters as Eqs. (\ref{recurrece}--\ref{entropy}) suggest. Observing that $S$ is null when computed for sufficient large or small $\varepsilon$, due to the absence of diversity of \textbf{RM} for both cases the vicinity  parameter is eliminated making use of the natural condition of maximum entropy  \cite{jaynes_1957}. Re-sampling the same time series in an  interval of $\tau$, the dependency on $\tau$ is also removed making use of the same maximum entropy principle.  Finally, after the two maximization processes, we get  
\begin{equation}
    S \equiv \mathsf{max}_{(\tau)}\left\{\mathsf{max}_{(\varepsilon)} \left [S(\varepsilon,\tau)\right ]\right\},
    \label{max_entropy}
\end{equation}  
an automatically set  and parameter-free quantity that can be computed for any time series.
The new  method brings advantages to data analysis, including the possibility to be used in artificial intelligence methods \cite{lopes_2020,prado_2020} . 

To illustrate how the time delay is automatically obtained by the  repetitive use of maximum recurrence entropy, we apply the method to the deterministic and chaotic Lorenz dynamical system \cite{Alligood1996chaos}, to a set of correlated and uncorrelated noise, \cite{lopes_2020}, and to an experimental data from an inductorless Chua's circuit \cite{torres_2000}.  

The Lorenz equations come from a reduction from seven to three differential equations, originally developed to model a convection motion in the atmosphere \cite{Alligood1996chaos}, and since then, a paradigmatic chaotic dynamical system described by 
\begin{equation}
       \dot{X}=\sigma (Y-X),\, \dot{Y}=X(\rho-Z)-Y,\,\mathrm{and}\, \dot{Z}=XY-bZ,
     \label{LorenzEquations}
\end{equation}
where $\rho$ and  $\sigma$ are the Rayleigh and the Prandtl numbers, and the quantity $b$ is a constant determined by the geometry of the problem. Here we fix $\rho=28$,  $\sigma=10$ and  $b=8/3$, a standard set of parameter for chaotic behavior of the Lorenz system \cite{Alligood1996chaos}.
\begin{figure}[!htpb] 
\centering\includegraphics[width=1.0\columnwidth]{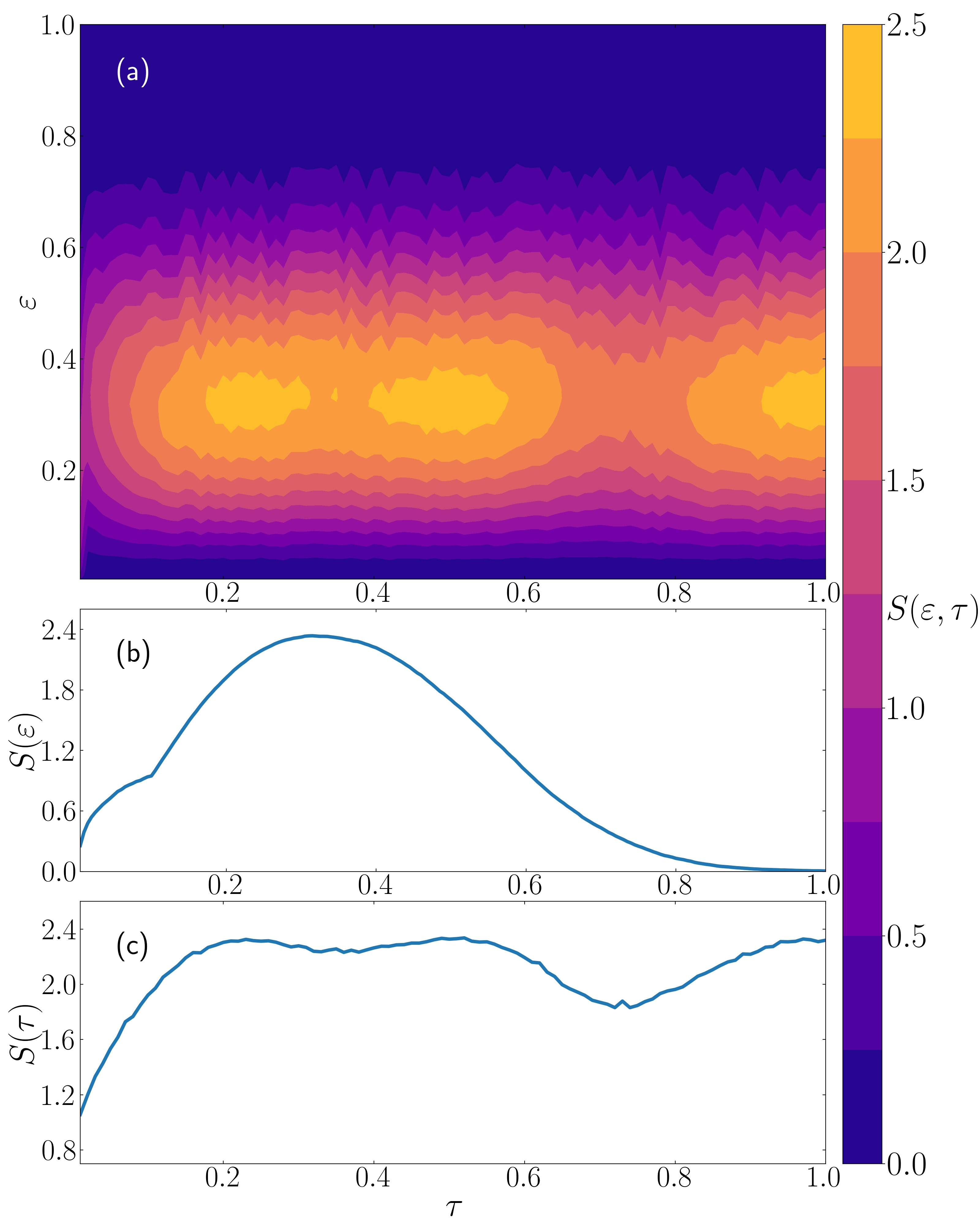}
\caption{(a) Contour plot of the recurrence entropy $S(\varepsilon,\tau)$,  using $N=2$ for the Lorenz system. 
The three maximum values of the recurrence entropy are $2.32$, $2.34$ and $2.33$, respectively. (b) $S(\tau)\equiv \mathsf{max}_{(\varepsilon)} [S(\varepsilon,\tau)]$. (c) $S(\varepsilon)\equiv \mathsf{max}_{(\tau)} [S(\varepsilon,\tau)]$.  The smallest value of $\tau$ for which $(S)$ can be computed gives the suitable time delay for the time series, however other values of $\tau$ can also lead to suitable time delays.  \label{fig:MaxLorenz}
}
\end{figure}

Figure \ref{fig:MaxLorenz}(a) depicts color-coded contour plot for the amplitudes of $S(\varepsilon,\tau)$ using $N=2$ (similar results hold for other values of $N$) for the bi-dimensional parameter space ($\varepsilon\times\tau$) considering the time series of $x\equiv \sqrt{X^2+Y^2+Z^2}$ of the  chaotic Lorenz system. Results for $X, Y,$ and $Z$ Lorenz components are shown and discussed in the supplementary material. We have computed  $S(\varepsilon,\tau)$ based on time series of size $50\,000\,000$, integrated in time-steps of $0.01$ (final time of $500\,000$). Segments of $10\,000$ points are then re-sampled from the time series using the time-delay in the interval $(0.01,\,1.00]$. Profiles of  $S(\varepsilon)\equiv \mathsf{max}_{(\tau)} [S(\varepsilon,\tau)]$ and $S(\tau)\equiv \mathsf{max}_{(\varepsilon)} [S(\varepsilon,\tau)]$ are plotted in Figs. \ref{fig:MaxLorenz} (b) and (c). Fig. \ref{fig:MaxLorenz}(b) shows just one maximum as a function of $\varepsilon$, occurring for $\varepsilon\approx 0.315$ for almost the entire interval of $\tau$. When we consider the behavior of $S(\tau)$ in the delay time interval $0<\tau<2$ [a.u], Fig. \ref{fig:MaxLorenz}(c) shows a sequence of maxima. The first one occurs for $\tau\approx 0.23$ [a. u.] followed by a secondary maximum at $\tau\approx 0.52$ [a. u.]. The presence of two non multiple values of $\tau$ reflects the Lorenz dynamics shifting between the two lobes of the attractor and denounces the two distinct time scales of the dynamics. Results for other dynamical systems may be found in the supplementary material. Based on these characteristics of $S(\varepsilon,\tau)$ we state that:

\textbf{Conjecture 1.\,}\textit{If $S$ exists in a time series, then a suitable time delay for the time series is the value of the smallest value of $\tau$ for which $S$ can be computed.}

\begin{figure}[!htpb] 
\centering\includegraphics[width=1.0\columnwidth]{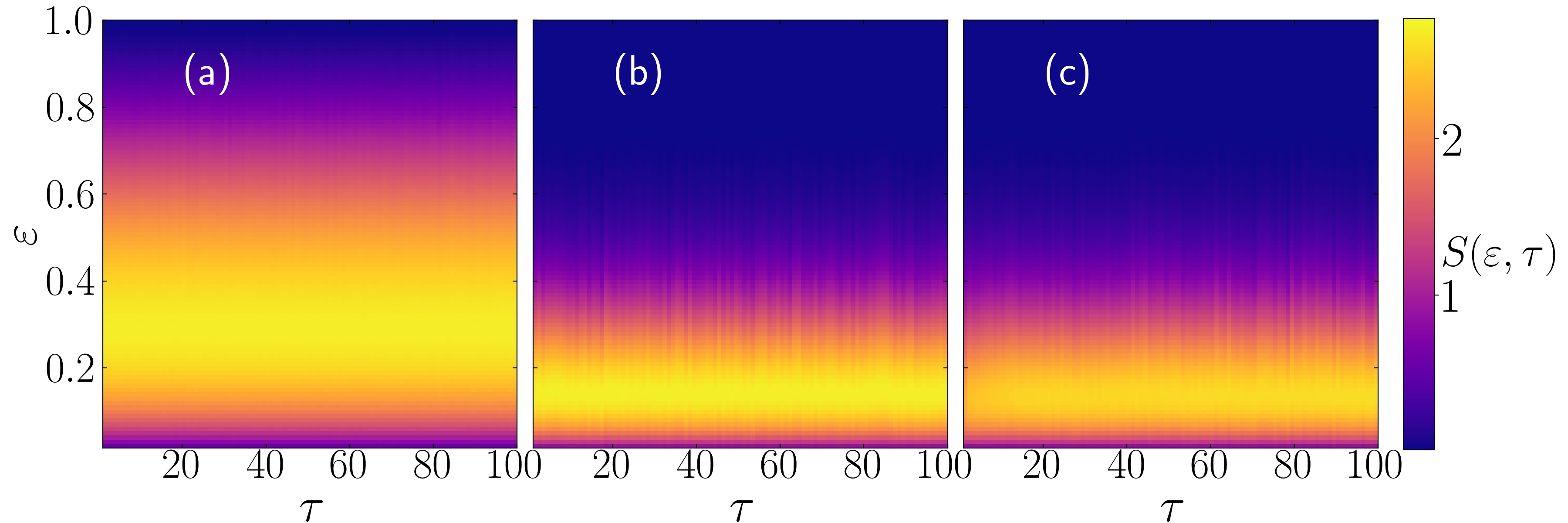}
\caption{(a) Contour plot of the recurrence entropy $S(\varepsilon,\tau)$,  using $N=2$ and computed for noisy data. (a) Uniform white noise. (b) Gaussian white noise. (c)  Correlated noise with power frequency spectrum following $1/f^\beta,\, \beta =1.0$ (c). Mean values of $S(\varepsilon ,\tau)$ result from  $10$ simulations for each pair    ($\epsilon,\,\tau$).  \label{fig:Maxnoise}
}
\end{figure}

Figs.  \ref{fig:Maxnoise} (a), (b), (c) depict results of  $S(\varepsilon,\tau)$ for uncorrelated (a,b) and correlated (c) stochastic signals. In these cases, we have considered time series of size $2\times 10^{5}$. As expected $S\equiv \mathsf{max}_{(\tau)}\left\{\mathsf{max}_{(\varepsilon)} \left [S(\varepsilon,\tau)\right ]\right\}$ can not be uniquely defined  as a function of $\tau$ and we state that:

\textbf{Conjecture 2.\,}\textit{
If $S$ does not exist in a time series, ($S$ can not be uniquely defined) the time series behaves like a stochastic sequence.}

For the case in which $S$ does not exist, $S(\tau)\equiv\mathsf{max}_{(\varepsilon)}  [S(\varepsilon,\tau)]\approx \mathrm{Constant}$ still distinguishes the level of correlation of the stochastic signal since distinct correlations lead to distinct maxima of $S(\varepsilon)$, see light yellow tones in Figs. \ref{fig:MaxLorenz}(d), (e), and (f) \cite{lopes_2020}.

Since suggested \cite{fraser_1986}, the time delayed mutual information $M(\tau)$ of a time series has been extensively used to determine the optimal time delay \cite{abarbanel_1993, abarbanel_2012}, $\tau_{\mathrm{optimum}}$, obtained as the first minimum of $M(\tau)$ as defined in Eq. \ref{mutual_info1}. 

\begin{figure}[!htpb]
    \centering
    \includegraphics[width=1.0\columnwidth]{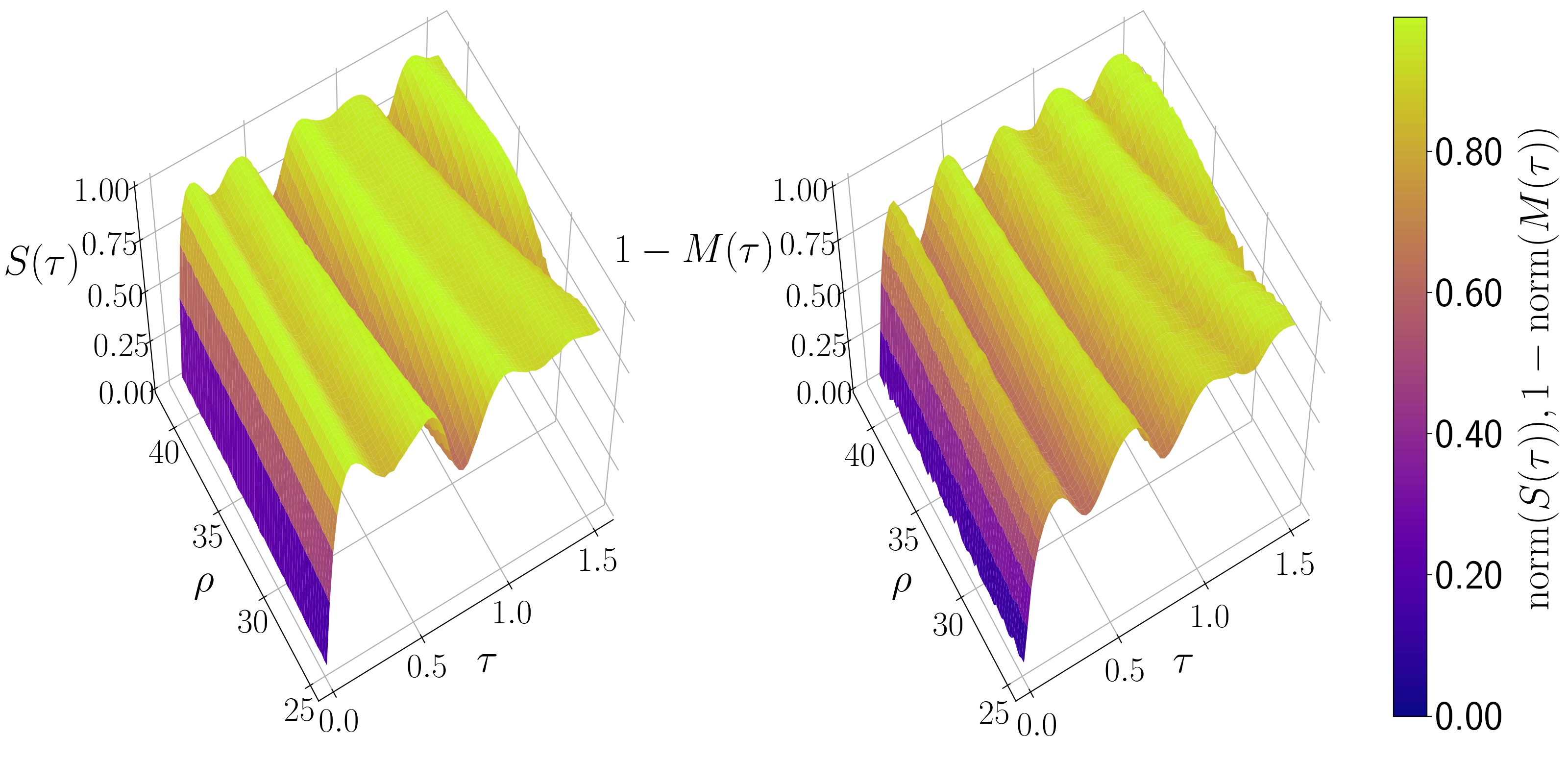}
    \caption{Comparative behavior of normalized  $S(\tau)$ given by Eq. \ref{max_entropy} (a) and $1-M(\tau)$, Eq. \ref{mutual_info1} (b) against the parameter $\rho$ of the Lorenz system. Both methods show a well defined picture of maxima where suitable $\tau$ can be chosen.}
    \label{fig:Mutual_Max_Lorenz}
\end{figure}

To show that the optimum time delay computed using $S$ is comparable to the value obtained using time delayed mutual information $M$, Figs. \ref{fig:Mutual_Max_Lorenz}(a, \,b) plot results of the normalized $[0,\,1]$ recurrence entropy $S$ (a) and the time delayed mutual information $M$ (b) computed  using Eqs. \ref{mutual_info1} and \ref{mutual_info2} and evaluating $RR$ using the traditional method as decribed in \cite{Marco_2003,marwan_2007} (we plot $(1-M)$ for better comparisons), for the Lorenz system and the interval $21<\rho<44$ of the Rayleigh number. Observe that maximum values of $S(\tau)$ and minimum values of $M(\tau)$ are completely equivalent for the entire interval of $\rho$. In particular, both  graphs depict large secondary possible time-delays, a natural behavior due to the pseudo-periodicity of the trajectory. 

To illustrate our methodology works for an experimental data, we use the inductionless Chua' circuit, which is a paradigmatic experimental setup  to study chaotic behavior \cite{torres_2000,chua_1993}. Circuit scheme and parameters for the experimental setup are given in supplementary material. The frequency of data acquisition is 25 kHz. 
\begin{figure}[!htpb]
    \centering
    \includegraphics[width=0.8\columnwidth]{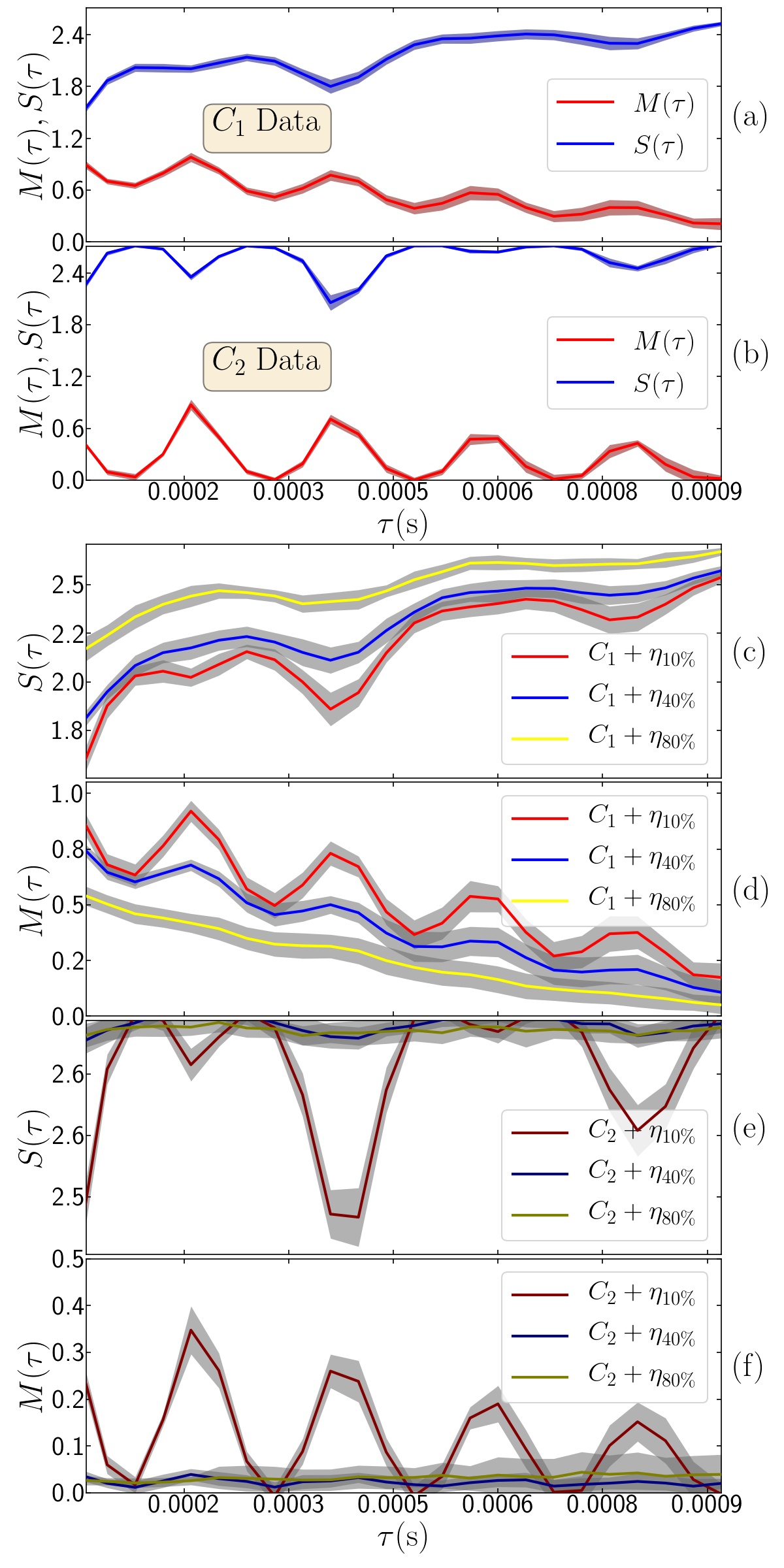}
    \caption{The recurrence entropy $S(\tau)$ and the time delayed mutual information function $M(\tau)$ for signals captured on both capacitor terminals of the Chua' circuit (panels a, b). The same results of panels a and b supposing distinct amounts of additive noise in the captured time series of both capacitors (panels c -- f). Color shadows are standard deviation considering 20 computation of $S(\tau)$ and $M(\tau)$ for each time delay $\tau$. The size of the time-delayed-reconstructed time series is 500 points.}
    \label{chua_data}
\end{figure}

Figures \ref{chua_data}(a,b) display results for $S(\tau)$ (blue curve ) and $M(\tau$) (red curve -- as computed in  \cite{Marco_2003}) based on the raw data collected on the terminals of the capacitors $C_1$ and $C_2$.  We have computed  $S(\tau)$ and $M(\tau)$ based on time series of size $100\,000$ corresponding to $4$ s of the circuit operation. Time series of $500$ points are then re-sampled from the raw data using a time-delay in the interval $(4\times 10^{-5}\mathrm{s} ,1\times 10^{-3}\mathrm{s}]$. Similarly to our results so far,  $S(\tau)$ depicts maxima at specific time delays which may be used to capture the suitable properties of the stationary state of the system (in this case the attractor properties). Clearly all extremes of $S(\tau)$ are in phase with the ones in the time delayed mutual information as discussed before.  Finally, we have also performed a noise addition tolerance analysis of our method, adding a gaussian non-time-correlated (white) noise with different amplitudes into the Chua' circuit data. Our results for $S(\tau)$ are depicted in Figs. \ref{chua_data}(c,e) corroborated by  results of $M(\tau)$ in Figs. \ref{chua_data}(d,f). For the case of noise  corrupted data obtained from $C_1$ terminals,  the sequence of suitable maximum values of  $S(\tau)$ are still present up to noise addition of amplitude of 80\%  of the Chua' maximum signal amplitude (yellow curve). The time delayed mutual information gives similar results till 40\% but for the extreme case of 80\% noise addition, it fails to display clear minimums. Using the raw data obtained form $C_2$ terminals, both methods give, again, similar results, but due to particular characteristics of the $C_2$ signal, $S(\tau)$ is able the recovery reliable maxima only for corrupted noise amplitude below 40\% of the maximum $C_2$ signal amplitude. 

At the end of our discussion, an important feature of our method must be pointed out. In order to get the results for the time delayed mutual information we have just time-shifted the time series and collected the results for $M(\tau)$,   in other words,  the maximum temporal resolution of the data has been used as Eqs. \ref{mutual_info1} and \ref{mutual_info2} show. To compute the results of $S(\tau)$ we have collect the data using a set of time delays, selecting the one for which the maximum value of the recurrence entropy is acquired, called here just $S$. So our method is applicable even when the temporal resolution is much lower than the ones used here. Tests using $M$ and more modest time resolutions show inconclusive results. 

In conclusion, we have shown that an information entropy based on recurrence microstates (\textbf{RM}) configures a reliable parameter-free quantifier to compute adequately time delay for collected data. The main advantage of the new methodology is the fact that it makes use of a physical principle, namely the repetitive maximization process of the recurrence and time-delay parameters of the microstate recurrence entropy, allowing the automatic setting of the appropriate time-delay of a time series of a dynamical system or an experimental data, exemplified here by the Chua' circuit, an electronic circuit used to obtain a chaotic signal experimentally. Due to the automated parameter selection process, we claim that it is a parameter-free theory and may be implemented in artificial intelligence methods to recovering correct information embedded in time series, rationalizing the process of data acquisition. We also show that the insertion of noise into time series, up to 40 \% of the time series amplitude  does not destroy the capability to predict the appropriated time delay of the signal.  Finally our method allows modest time resolution of the data, distinctly of other methods that make use of a high resolution data collection to infer the  appropriate time delay.  

We thank the Coordena\c c\~ao de Aperfeiçoamento de Pessoal de N\'{\i}vel Superior - Brasil (CAPES) - Finance Code 001,  the Conselho Nacional de Desenvolvimento Cient\'{\i}fico e Tecnol\'ogico,  CNPq - Brazil, grant numbers 302785/2017-5, 307907/2019-8, and 308621/2019-0
 and Financiadora de Estudos e Projetos (FINEP) for finantial support.
 \bibliography{biblio}
\end{document}